\documentstyle[preprint,aps]{revtex}
\begin{document}
\draft
\title
{BLACK HOLE FORMATION AND SPACE-TIME FLUCTUATIONS IN
TWO DIMENSIONAL DILATON GRAVITY
AND COMPLEMENTARITY}
\author{Sumit R. Das \footnote{E-mail: das@theory.tifr.res.in}}
\address
{Tata Institute of Fundamental Research \\
Homi Bhabha Road , Bombay 400005, INDIA}
\author{Sudipta Mukherji \footnote{E-mail: mukherji@ictp.trieste.it}}
\address
{International Centre for Theoretical Physics \\
I-34100 Trieste. ITALY}
\maketitle

\begin{abstract}
We study black hole formation in a model of two dimensional
dilaton gravity and $24$ massless scalar fields with a boundary. We find
the most general boundary condition consistent with perfect reflection of
matter and the constraints. We show that in the semiclassical
approximation  and for the generic value of a parameter which
characterizes the boundary conditions, the boundary starts receeding to
infinity at the speed of light whenever the {\em total} energy of the
incoming matter flux exceeds a certain critical value.  This is also the
critical energy which marks the onset of black hole formation. We then
compute the quantum fluctuations of the boundary and of the rescaled
scalar curvature and show that as soon as the incoming energy exceeds this
critical value, an asymptotic observer using normal time resolutions will
always measure large quantum fluctuations of space-time near the {\em
horizon}, even though the freely falling observer does not. This is an
aspect of black hole complementarity relating directly to quantum gravity
effects.
\end{abstract}
\narrowtext
\newpage

\def\be{\begin{equation}}
\def\ee{\end{equation}}
\def\ben{\begin{equation}}
\def\een{\end{equation}}
\def\bbb{\bibitem}
\def\bea{\begin{eqnarray}}
\def\eea{\end{eqnarray}}
\def\nn{\nonumber}

\def\half{{1 \over 2}}
\def\gab{g_{ab}}
\def\sqg{{\sqrt{g}}}
\def\inuv{\int du dv}
\def\pu{\partial_u}
\def\pv{\partial_v}
\def\klog{{\rm{log}}}
\def\inu{\int du}
\def\inv{\int dv}
\def\gab{g_{ab}}
\def\sqrtg{{\sqrt{g}}}
\def\intuv{\int du dv}
\def\pu{\partial_u}
\def\pv{\partial_v}
\def\pvp{\partial_{v'}}
\def\klog{{\rm{log}}}
\def\inu{\int du}
\def\inv{\int dv}
\def\xp{X^+}
\def\xm{X^-}
\def\xtp{Y^+}
\def\xtm{Y^-}
\def\gp{g^+(u)}
\def\gm{g^-(v)}
\def\hm{h_-}
\def\sqd{{\sqrt{\Delta}}}
\def\sqdm{{\sqrt{-\Delta}}}
\def\zp{z^+}
\def\zm{z^-}
\def\delb{{{\sqrt{\Delta}} \over \beta}}
\def\ksinh{{\rm{sinh}}}
\def\ep{\epsilon}
\def\xpcl{X^+_{cl}}
\def\xpq{X^+_{qu}}
\def\xtmcl{Y^-_{cl}}
\def\xtmq{Y^-_{qu}}
\def\ficl{f^i_{cl}}
\def\fiq{f^i_{qu}}
\def\spi{\sqrt{2\pi}}
\def\ypl{{y^+ \over L}}
\def\yml{{y^- \over L}}
\def\yppl{{y^{+'} \over L}}
\def\ympl{{y^{-'} \over L}}
\def\ltr{L^{tr}}
\def\tf{{\tilde f}^i}
\def\keto{|0,in>}
\def\brao{<0,in|}
\def\curv{{\tilde R}}
\def\pp{p^+}
\def\tf{{\tilde{f}}}
\def\tL{{\tilde L}}
\def\yp{y^+}
\def\ym{y^-}
\def\ypp{y^{+ '}}
\def\ymp{y^{- '}}
\def\pyp{\partial_{y^+}}
\def\pym{\partial_{y^-}}
\def\hk{{\hat{\kappa}}}

\def\cmp{{\em Comm. Math. Phys.}}
\def\pr{{\em Physical Review,}}
\def\mpla{{\em Modern Physics Letters,}}
\def\nph{{\em Nuclear Physics,}}

\section{Introduction}

Recently, the question of information loss during the process of black
hole evaporation \cite{INFO} have been intensively studied \cite{REV}. The
reason behind this renewed interest is the discovery of black hole
solutions in two dimensonal dilaton gravity models \cite{MSW} and exact
conformal field theories describing two dimensional black holes \cite{WIT}
and the demonstration that black hole formation and subsequent evaporation
can be {\em in principle} studied in two dimensional models
\cite{CGHS}.

Much of the work in the past two years have dealt with the problem in the
semiclassical approximation \cite{CGHS},
\cite{BANKS,RST,BGHS,HAW,STROM,BILAL,DEALWIS,GIDST,PARK,BANK,LOWE,PIRAN,MARTIN}
in which the dilaton and graviton fields are treated as classical
quantitites wheras the matter fields are treated quantum mechanically - an
approximation which is valid if the number of matter fields $N$ is large
with the product $N e^\phi$ kept finite ($\phi$ denotes the dilaton
field).  A more definitive analysis can be carried out for
Reissner-Nordstrom black holes \cite{TRIV}.  In most models the
semiclassical approximation is valid in a region of space time bounded by
a critical line where $\phi(x)$ becomes equal to a critical value
$\phi_c(x)$. Nevertheless, the evolution of the system away from this
boundary can be studied analytically for some models or numerically for
others. The results of the semiclassical picture seems to converge on a
picture where information is lost from the observable part of the
universe.

It has been argued, however - most persuasively by 't Hooft - that one
cannot really ignore quantum fluctuations of gravity even if the black
hole is of large mass and $N$ is large since high energy physics near the
horizon appear as normal energy processes to the aymptotic observer due to
the very large redshift.

A possible approach to an exact quantum treatment of black holes comes
from the connection of matrix models with static black hole backgrounds,
viz. a certain integral transform of the collective field of the matrix
model behaves as a "tachyon" field in the black hole background of the two
dimensional critical string.
\cite{BHMM}. One remarkable result is that the singularity of
the background disappears in the full nonperturbative answer
\cite{DMWBH} . In
fact the same holds true for the exact one particle wave function of the
problem as well as for more general incoming waves \cite{WAVE} .  A
different matrix model black hole connection has been proposed in
\cite{JEVYON} .
We expect the above conclusions to hold in this model as well
\footnote{For yet another proposal for description of black holes in matrix
models see \cite{RUSSOA} }.  However, this connection is as yet known only
for {\em static} black holes and one does not have a description of black
hole formation.  Furthermore, in matrix models gravity and dilaton fields
do not appear in the action explicitly : in a sense they are integrated
out and their quantum effects are completely contained in the collective
field theory or the fermionic field theory. It is useful to have models
where one can study gravity fluctuations explicitly and exactly and infer
how they affect the quantum evolution of black holes.

In a recent paper one such model with $24$ scalar fields in the presence
of two dimensional dilaton gravity with a boundary has been studied by
Verlinde and Verlinde and by Schoutens,Verlinde and Verlinde
\cite{VV} . In
a fiducial coordinate system with light-cone coordinates $(u,v)$ the
boundary is taken to be $u=v$. However, in terms of physical coordinates
which become minkowskian far away, the boundary is dynamically determined
by the infalling matter and undergoes quantum fluctuations. In two
dimensions there is no physical local degree of freedom of the
graviton-dilaton system, but the degree of freedom corresponding to a
dynamical boundary survives. Thus the effect of boundary fluctuations in
this model serves as a useful toy model for studying effects of quantum
gravity fluctuations. In fact this model has some features of the moving
mirror problem \cite{MIRR} .

In this paper we study this model further with a view to understand
quantum gravity effects. We first find the most general boundary
conditions which leads to a perfect reflection of the matter energy
momentum tensor and consistent with the constraints.  These boundary
conditions are parametrized by a single parameter $\beta$. The boundary
conditions of \cite{VV} correspond to $\beta = 0$. In a recent paper
\cite{HVER} similar boundary conditions have been used to study the
dynamics of the boundary as a dynamical moving mirror. Our discussion is
complementary to this paper and concentrates on the space-time features of
the model.

The physics of the model is very different for $\beta = 0$ and $\beta \ne
0$. For $\beta = 0$ there is no meaningful classical solution. There is a
semiclassical solution in which the boundary "jumps forward" and becomes
spacelike as soon as the incident matter {\em energy density} exceeds a
critical value. At the same time a black hole singularity forms which
meets the apparent horizon at a finite time, like the other semiclassical
models studied in the literature.

On the contrary the $\beta \ne 0$ model makes sense classically.  The
boundary now receedes away as the matter impinges on it. When the {\em
total} energy of the incoming matter exceeds a critical value, the
boundary starts receeding with the speed of light and with unbounded
acceleration at a time much {\em later} than the time of impact. In the
classical theory this is also the critical value beyond which a black hole
singularity forms. At the semiclassical level, the critical value of the
energy has a different dependence on $\beta$. The solution is, in fact,
very similar to that in the RST model \cite{RST} with a similar picture of
black hole evaporation.

Finally we quantize the $\beta \ne 0$ model fully by expanding around the
classical solution representing black hole formation.  We calculate some
quantum fluctuations of quantities related to space-time in this theory,
like the dispersion of the line element on the boundary and the dispersion
of the rescaled scalar curvature.  As in any field theory these
dispersions depend on the ultraviolet cutoff which simply express the fact
that one generically sees more and more quantum fluctuation as one goes to
shorter and shorter distances.  In other words, the dispersion depends on
on the resolution time of the measuring apparatus.  If all observers use a
resolution time of order one (in natural units) we show that an infalling
observer finds a finite dispersion. An asymptotic observer, however, has a
rather different conclusion. So long as the incoming energy is below the
critical value, the aymptotic observer also finds a finite dispersion with
a finite resolution time. However as soon as the energy exceeds a critical
value and a horizon is formed the dispersion grows as one approaches the
horizon and becomes infinite at the horizon - even with a finite
resolution time. Thus the aymptotic observer always finds that the quantum
fluctuations of the dilaton and metric diverge at the horizon, which may
be quite far from the singularity. The result is quite consistent with the
ideas put forward by 't Hooft \cite{THOFT}and with ideas of black hole
complementarity
\cite{SUSSTH} . The phenomenon
is quite similar to the infinite spreading of strings moving in black hole
or rindler spacetimes considered by Susskind
\cite{SUSS}.
It is unclear whether the presence of these large fluctuations of space
time as observed by an asymptotic observer signifies the invalidity of the
semiclassical approximation even for large black holes.

\section{ The Model and Boundary conditions}

The model consists of a space-time metric in two dimensions $\gab(u,v)$
and a dilaton field $\phi(u,v)$ and $24$ massless scalar fields $f^i(u,v)$
with the action
\bea
S & = & {1 \over 2\pi} \inuv~\sqg[e^{-2\phi}(R - 4(\nabla \phi)^2 \nn \\
- 4 \lambda^2) + \half (\nabla f^i)^2] \label{eq:one}
\eea
Here $R$ is the scalar curvature and $\lambda^2$ is a cosmological
constant which can be set to one by a choice of scale.  The space-time has
a boundary, which is the fixed  line $u = v$ in a fiducial set of null
coordinates $(u,v)$. These are defined in terms of the space coordinate
$x$ and the time coordinate $t$ by $u = t + x$ and $v=t-x$.

We will work in the conformal gauge $\gab = e^{2\rho}~\eta_{ab}$. Let us
introduce two free chiral fields $\xp(u)$ and $\xm(v)$. These are related
to the metric and dilaton fields by
\ben
 e^{2(\rho - \phi)} = \pu \xp (u)
\pv \xm (v) \label{eq:bcone}
\een
The remaining equation of motion is then
\ben
\pu \pv e^{-2\phi} = - \pu \xp (u) \pv \xm (v) \label{eq:two}
\een
 and constraints simply state the total energy momentum tensor vanishes
\ben
T^g_{uu} + T_{uu}
= 0~~~~~~~~T^g_{vv} + T_{vv} = 0 \label{eq:three}
\een
 Here $T^g_{uu},
T^g_{vv}$ stand for the gravity-dilaton part of the eenergy momentum
tensor while $T_{uu}, T_{vv}$ the matter part. We will often omit the
subscripts since they are chiral. Since the matter is conformally coupled
we may write
\ben
f^i(u,v) = f^{i+}(u) - f^{i-}(v) \label{eq:four}
\een

The general solution of (\ref{eq:two}) may be written as
\ben
 e^{-2\phi} = -\xp (u)
\xm (v) + g^+ (u) + g^- (v) + K \label{eq:bctwo}
\een
where the functions $g^\pm$ have to be determined by solving the
constraints (\ref{eq:three}),
and $K$ is an integration constant.  The expressions
for $T^g$ become
\bea
&T_{uu}^g = -\pu
\gp {\pu^2 \xp \over \pu \xp} + \pu^2 \gp  \nn \\
 & T_{vv}^g = -\pv \gm
{\pv^2 \xm \over \pv \xm} + \pv^2 \gm \label{eq:bcfour}
\eea
 Now define two
new fields $\xtp (v)$ and $\xtm (u)$ by
\bea
& \pu \gp = \xtm (u) \pu \xp(u) \nn \\
& \pv \gm = \xtp (v) \pv \xm (v)
\label{eq:bcfive}
\eea
One then has
\ben
 T_{uu}^g = \pu \xtm (u) \pu \xp (u)~~~~~~T_{vv}^g = \pv
\xtp (v) \pv
\xm (v) \label{eq:bcsix}
\een

Finally we write down the expression for the scalar curvature in terms of
the fields introduced above. This is given by
\bea
 R & = & 8 e^{-2\rho} \pu \pv \rho  \nn \\
& = & 4[1 + e^{2\phi}(\xm (v) - \xtm (u))(\xp (u) - \xtp
(v))]\label{eq:bcdone}
\eea
 We also introduce the "rescaled" scalar
curvature
\ben
{\tilde R} \equiv e^{-2\phi}~(R-4) \label{eq:bcdtwo}
\een
which will be useful in later calculations. In fact this is the quantity
that appears in the action.  The expression (\ref{eq:bcdone})
shows that the zeroes
of $e^{-2\phi}$ would be generically curvature singularities, unless the
expression which multiplies it vanishes.  Furthermore any region where any
of the fields $X^\pm, Y^\pm$ diverge is also potentially a region of
curvature singularity.

\subsection {Boundary Conditions}

The boundary conditions we want to impose are (i) The dilaton field must
be constant along the boundary, i.e. $(\pu + \pv)e^{-2\phi} = 0$ along
$u=v$ and (ii) the matter must be perfectly reflected at the boundary.
This sets $f^{i+}(u) = f^{i-}(u)$ and ensures that $T_{uu}(u) = T_{vv}(v)$
at $u=v$. Since the total energy momentum tensor must vanish (or a
constant in the quantum theory) we must, for consistency, also require
that (iii) the energy momentum tensor of gravity is also reflected off the
boundary perfectly, i.e. $T_{uu}^g(u) = T_{vv}^g(v)$ at $u=v$.

The conditions on the dilaton and metric fields are nontrivial.  The
condition (i) is
\ben
[ \pu \xp(\xtm (u) - \xm (v)) + \pv \xm (\xtp (v) - \xp (u))] =
0 \label{eq:bcseven}
\een
at $u = v$.
 In \cite{VV} each term
in (\ref{eq:bcseven}) was separately set to zero,
which also automatically ensured
(iii). This sets both $\pu e^{-2\phi}$ and $\pv e^{-2\phi}$ to zero at the
boundary and corresponds to the boundary conditions used in \cite{RST}.
Clearly this is not the general condition.

If there was no boundary we could have used the remaining
reparametrization invariances to choose coordinates such that $\xp(u)$ and
$\xm (v)$ are both of a desired form. However the boundary conditions
relate left moving and right moving modes, which means that the remaining
reparametrizations of $u$ and $v$ are not independent. We can, however,
fix one of the fields $X^\pm$. In this paper we will often use a gauge
such that $\xp (u) = u$. We will see later that this choice corresponds to
a choice of Kruskal coordinates. The form of $\xm(v)$ has to be now
determined by solving the constraints as we will show.

Let us introduce a new field $\hm (v)$ defined as
\ben
 \pv \xm (v) =  \hm^2 (v) \label{eq:bceight}
\een
The most general form for $\xtp,\xtm$ which satisfies the condition
(\ref{eq:bcseven})  may be easily seen to be, in the gauge $\xp (u) = u$
\bea
 \xtm (u) & = & \xm (u) + F[u,\hm (u)] \nn \\
\xtp (v) & = & v - F[v,\hm (v)] {1 \over \hm^2}\label{eq:bcnine}
\eea
where $F[x,\hm(x)]$ is a general function of $x$ and a functional of $\hm
(x)$. Now substitute (\ref{eq:bcnine})
into the expressions (\ref{eq:bcsix}) and get
\bea
T_{uu}^g & = & \hm^2 + {dF \over du} \nn \\
T_{vv}^g & = & \hm^2 - {dF \over dv} + {2 F \pv \hm \over
\hm}\label{eq:bcten}
\eea
Requiring these two expression to match at $u = v$ we get the condition
\ben
 { \partial F \over \partial u} + { \delta F \over \delta \hm} \pu \hm
 = { F \pu \hm \over \hm}\label{eq:bcelevena}
\een
 whose unique solution is
\ben
 F[u,\hm(u)] = \beta \hm (u) \label{eq:bceleven}
\een
where $\beta$ is an integration constant.  The boundary conditions in
\cite{VV} corresponds to $\beta =0$.

The boundary conditions (\ref{eq:bcnine})
may be considered as operator conditions
in the quantum theory. In the semiclassical theory, however, the
expression for the energy momentum tensor has additional terms coming from
the conformal anomaly and the form of the solution for $F[u,\hm(u)]$ are
correspondingly different. This will be discussed in a later section.

\section{The Classical Solution}

Given some energy momentum tensor of incoming matter the classical
solution is obtained by solving the equations $T^g(u)+T(u)=0$ and $T^g(v)
+ T(v) =0$ where $T^g$ is obtained from (\ref{eq:bcten}) with $F$ given by
(\ref{eq:bceleven}).
These two equations are in fact identical (since the reflection
condition on matter fields set $T(u)=T(v)$) and become
\ben
 \hm^2 + \beta \pu \hm + T(u) = 0 \label{eq:bctwelve}
\een
Given a solution of $\hm (u)$ it is straightforward to solve for $\xm
(u)$. This is given by
\ben
\xm = \int^u  \hm^2 + C
\een
where $C$ is an integration constant. This becomes, after using
(\ref{eq:bctwelve}),
\ben
 \xm (u) = -\int^u T(u) - \beta \hm + C \label{eq:bcthirteen}
\een

The gravity and dilaton fields may be now readily obtained. From the
definition of $g^\pm$ and $\xtp, \xtm$ in (\ref{eq:bctwo})
and (\ref{eq:bcfive}), the
relations (\ref{eq:bcnine}) and (\ref{eq:bceleven})
and the equation (\ref{eq:bctwelve})
we get
\bea
 g^+(u) & = & u \xm (u) + \beta u \hm (u) - \int^u u' T(u') \nn \\
g^- (v) & = & -\beta v \hm(v) - \int^v dv'v' T(v')\label{eq:bcseventeen}
\eea
Putting these together we get
\bea
e^{-2\phi} & = & u[\xm (u) - \xm (v)] + \nn \\
& & [\int^u du'u' T(u') - \int^v dv'
v' T(v')] \nn \\
&  & + \beta[u \hm (u) - v \hm (v)] + K \label{eq:bceighteen}
\eea
where $K$ is an integration constant, which is the value of $e^{-2\phi}$
on the boundary.

Note that in this classical problem, $\beta$ can be scaled out of the
problem by rescaling $\hm (u) \rightarrow \beta\hm (u)$ and $T(u)
\rightarrow \beta^2 T(u)$ in (\ref{eq:bctwelve}).
This leads to a scaling of $\xm
\rightarrow \beta^2 \xm$. This immediately shows that $\beta = 0$ is a
rather singular limit. In the following we will work mostly with $\beta
\ne 0$ unless otherwise stated.

\subsection{Solution with Shock wave matter}

We now find the classical solution in the presence incoming matter in the
form of a shock wave whose matter energy momentum tensor is given by
$T(u)={1 \over u^2}T\delta(u-1)$ (we have chosen the location of the shock
wave at $u=1$ by a suitable shift of coordinates). We will solve
(\ref{eq:bctwelve}) with the
condition that the spacetime is flat linear dilaton
vacuum before the shock wave arrives (i.e. in the region $u < 1$ for all
$v$). With this condition the solution to (\ref{eq:bctwelve})
can be found in all
of the $u-v$ space :
\ben
\hm (v)  = {\beta \over v}[1 - ({a \over a + v (1-a)})\theta (v-1)]
\label{eq:nine}
\een
where $a = {T \over \beta^2}$.  This is an explicitly real solution and
since $\pv \xm (v) = \hm^2 (v)$ it may appear that $\xm (v)$ is a
monotonically increasing function, so that the boundary is everywhere
timelike. However, (\ref{eq:nine})
clearly shows that $\hm (v)$ blows up for some
positive value of $v$ whenever $a >1$.  This becomes clear when one looks
at the solution for $\xm (v)$ obtained by plugging in (\ref{eq:nine}) into
(\ref{eq:bcthirteen})
\ben
\xm (v) = \beta^2[(-{1 \over v} + a)\theta(1-v) -({1 - a \over a + (1-a)
v}) \theta(v-1)] \label{eq:ten}
\een
 which shows that $\xm (v)$ blows up at
$v = {a \over a-1}$ when $a > 1$. The solution for the dilaton field is
given by
\bea
e^{-2\phi} & = & \beta^2 {u \over v}~~~~~~~~~~~u,v < 1 \nn \\
& = & \beta^2(u({1 \over v} - a) + a) ~~~~~~~~~~u > 1, v < 1 \nn \\
& = & \beta^2 {(a + (1-a)u) \over (a + (1-a)v)}~~~~~~~~~u,v > 1
\label{eq:eleven}
\eea
The meaning of the parameter $\beta$ is now clear. This is simply the
value of $e^{-\phi}$ at the boundary.  The metric is best expressed in
terms of the coordinates $z^\pm$ defined as $z^+ = u$ and $z^- = -{1\over
v}$. This becomes
\bea
ds^2 & = & {d\zp d\zm \over \zp \zm}~~~~~~~~~~u,v < 1 \nn \\
& = & {d\zp d\zm \over (a-\zp(\zm + a))} ~~~~~~~~~~u > 1, v < 1 \nn \\
& = & ({a\zm + (1-a) \over a\zm + (1-a) \zm \zp})~d\zp d\zm~~~~~~~~~u,v > 1
\label{eq:twelve}
\eea
 From (\ref{eq:twelve}) and (\ref{eq:ten}) it may be verified that the
asymptotically minkowskian coordinates are $\klog(\xp (u))$ and
$-\klog(-\xm(v))$. As in theories of gravity field variables become
physical coordinates. In the quantum theory these "coordinate fields"
become operators and have important consequences \cite{VV}.

The profile of the boundary in the $\xp, \xm$ coordinates is simply the
curve of $\xm (v)$ as a function of $v$. From the above expressions it is
clear that as long as $a < 1$, the boundary is always timelike.  At the
point where the pulse hits the boundary there is a discontinuity, but it
never has an unbounded acceleration.  Furthermore $\xm (v)$ always stays
negative, and no horizons are formed.  For $a > 1$, $\xm (v)$ turns
positive {\em before} the pulse hits the boundary. At the impact point,
there is a usual discontinuity, but nothing dramatic happens at ths point.
However at a later time the acceleration increases without bound, the
velocity approaches the speed of light and $\xm$ becomes infinite at a
point $v = {a \over a-1}$.  As $a$ increases this point approaches the
point $u=v=1$ where the shock have hits the boundary. It is not meaningful
to continue the solution beyond this point.  In fact the asymptotic
observer will see the boundary running away at the point where $\xm$ hits
a zero, since this is the end of the asymptotic coordinate system. This
instability has been observed in \cite{HVER} and is different from the
instability discussed above.  The latter occurs for a Kruskal observer,
using the $X^\pm$ coordinates.

Thus there is a critical value of the {\em total} energy of the incoming
pulse beyond which the boundary runs away. Note that we are dealing with a
delta function shock wave for which the {\em energy density} is always
unbounded.

The interesting point in this model is that it is precisely at this value
of the total energy that a black hole starts forming. This is clear from
the solution to the dilaton and the metric fields in (\ref{eq:eleven})
and (\ref{eq:twelve}).
Note that the conformal factor in the $z^\pm$ coordinates is exactly equal
to $e^{2\phi}$. From (\ref{eq:twelve}) we see
that there are no singularities in
the region before the incident pulse. In the region $u > 1, v < 1$ the
metric is exactly like the standard black hole metric. We thus have a
potential curvature singularity along the spacelike line $\zp(\zm + a)=a$.
Since this region in the $u,v$ space corresponds to $\zp > 1, \zm < -1$,
there is no singularity when $a < 1$. For $a > 1$ there is a singularity
which asymptotes to $\zm = -a$. The null line $\zm = -a$ which correspond
to $v = {1 \over a}$ is the event horizon which in this case also
coincides with the apparent horizon (defined in the usual way by $\pu \phi
= 0$). In the region $u,v > 1$ the singularity is along the null line $u =
{a \over a-1}$ which begins at the point where the singularity in the
region II intersects with the reflected wave.

The expression (\ref{eq:bcdone}) may be used to
compute the curvature scalar. The
result is that $R$ diverges on the part of the singularity in the $u >1, v
< 1$ region, but is finite along the null singularity in the $u,v > 1$
region. However, in this model the kinetic energy terms contain a factor
of $e^{-2\phi}$ and zeroes of $e^{2\phi}$ are genuine singularities even
when the curvature is finite.

The kruskal diagrams for the solution (in the coordinates $X^\pm$) are
shown in Fig. 1 for $a < 1$ and in Fig. 2 for $a > 1$. Note the horizon is
the line $\xm = 0$ while the reflected pulse is along $\xm = \beta^2
(a-1)$.

\section{The semiclassical solution}

We now outline what happens in the model when we take into account quantum
fluctuations of the matter fields, but still treat the dilaton gravity
sector classically. As is well known, this has several effects. First,
there is a constant negative vacuum energy due to normal ordering effects.
In the $(u,v)$ coordinates this vacuum energy (for $N$ scalar fields) is
simply $-{N \over 48 u^2}$ Secondly the conformal anomaly induces new
terms in the action which modifies the equations of motion as well as the
form of the energy momentum tensor for the dilaton gravity system.
However, as demonstrated in \cite{RST} one can choose counterterms such
that the dynamical equations of motion of the semiclassical theory are
similar to those of the classical theory with the replacement
\ben
e^{-2\phi} \rightarrow \Omega \equiv {1 \over \kappa}[e^{-2\phi} +
\kappa \phi]
\label{eq:thirteen}
\een
Here $\kappa = {N \over 24}$. Thus the solution to $\Omega$ is still given
by the general form in equation (\ref{eq:bctwo}).
With the same definition of the
fields $\xtm,\xtp$ as in (\ref{eq:bcfive})
the expressions for the gravity part of
the energy momentum tensor becomes, instead of (\ref{eq:bcsix})
\bea
T^g_{uu} & = & \kappa[{\bar T}^g_{uu} + {\hk \over 2 \kappa}
[(\pu \klog \pu
\xp)^2 - 2 \pu^2 \klog \pu \xp]] \nn \\
T^g_{vv} & = & \kappa[{\bar T}^g_{vv} + {\hk \over 2 \kappa}
[(\pv \klog \pv
\xm)^2 - 2 \pv^2 \klog \pv \xm]]\label{eq:sone}
\eea
where ${\bar T}^g$ denote the classical
contributions in (\ref{eq:bcsix}) and
$\hk = \kappa - 1 = {N - 24 \over 24}$.  The most general boundary
conditions may be now derived in the gauge $\xp (u) = u$ following the
same steps as in the previous sections. With the same ansatz as in
(\ref{eq:bcnine})
and the requirement that the gravity part of the energy momentum
tensor is perfectly reflected at the boundary one gets instead of
(\ref{eq:bcelevena}) the following equation for $F[\hm,u]$
\ben
{ \partial F \over \partial u} + { \delta F \over \delta \hm} \pu \hm
 = { F \pu \hm \over \hm} + \hk[(\pu \klog \hm)^2 - \pu^2 \klog
\hm]\label{eq:stwo }
\een
whose general solution is
\ben
F[\hm (u),u] = \beta \hm (u) - {\hk \over \kappa}\pu \klog \hm (u)
\label{eq:sthree}
\een
This shows that, in particular, the boundary conditions in \cite{VV},
which is identical to that used in \cite{RST} cannot be valid unless $N =
24$ \footnote{S. Trivedi has informed us that he and his collaborators
have also discovered that the boundary conditions $\partial_\pm \Omega =
0$ are not consistent in the RST model. \cite{TRP}}. The equation
satisfied by $\hm (u)$ now becomes
\ben
 \hm^2 + \beta \pu \hm - {\hk \over \kappa}
 \pu^2 \klog \hm + T(u) = {1 \over 2 u^2}
\label{eq:sthreea}
\een
In general these equations are difficult to solve exactly.

The semiclassical limit could be a good description when $N$ is very
large. However one may get some insight into quantum effects of matter for
the present case of $N=24$ where the semiclassical equations can still be
solved exactly. This is because in this case the anomaly is absent and the
only modifications are the inclusion of the vacuum energy and the
replacement of the dilaton field by $\Omega$ in (\ref{eq:thirteen}).

\subsection{Semiclassical solution with shock wave for $\beta \ne 0$}

To solve the constraints for a shock wave one has
to solve (\ref{eq:bctwelve}) with
\ben
T(u)={1 \over 2 u^2}[T\delta(u-1) - 1]\label{eq:thirteenh}
\een
We require that the solution to the left of the shock wave is the flat
minkowski space linear dilaton vacuum. This yields the following solution
\bea
\hm(u) & = & {1 \over 2 u}[(\beta +\sqd) \nn \\
& & -{2T\sqd \over (T+(\beta \sqd - T)u^{\delb})}\theta (u-1)]
\label{eq:fourteen}
\eea
 The solution for $\xm
(u)$ is
\bea
\xm(u) & = & -{1 \over u }[\gamma + Tu \theta(u-1) - \nn \\
& & {\beta T\sqd \over (T+(\beta \sqd - T)u^\delb)}\theta (u-1)]
\label{eq:fifteen}
\eea
where
\ben
\Delta = \beta^2 +2~~~~~~~\gamma = \half[1 + \beta (\beta +
\sqd)] \label{eq:sixteen}
\een
The field $\Omega(u,v)$ is then given by
\bea
\Omega(u,v) & = & \gamma {u \over v}
-\half \klog({\gamma u \over v})~~~~~~~~~~u,v < 1 \nn \\
& = & u({\gamma \over
v} - T ) + T -\half \klog({\gamma u \over v})~~~~~~~~u >1, v<1 \nn \\
& = & {\gamma u \over v} + [1 - {u \over v}]
{\beta T\sqd \over (T+(\beta \sqd - T)v^\delb)} \nn \\
&  & - \half \klog({\gamma u \over v})~~~~~~~~u,v > 1
\label{eq:seventeen}
\eea
In writing down (\ref{eq:seventeen})
 we have chosen an integration constant in the
solution for $\Omega(u,v)$ such that in the region before the incoming
wave, one has a standard linear dilaton vacuum with $e^{-2\phi} = {\gamma
u \over v}$. As in the classical solution the asymptotically minkowskian
coordinates are $\pm \klog(\pm X^\pm)$.

 From (\ref{eq:fifteen}) it is clear
that $\xm (u)$ diverges for some finite $u = u_0$ where
\ben
u_0 = [{T \over T - \beta\sqd}]^{{\beta \over \sqd}}
\een
 whenever $T > T_0 =
\beta\sqd $. Thus $\beta\sqd$ is the critical value of the incoming energy
beyond which the boundary runs away.

Before looking for singularities let us look for the presence of apparent
horizons. This is given by the curve along which $\pu \Omega = 0$ (since
$\pu \phi = 0$ implies $\pu \Omega = 0$).  It is trivial to see that there
is no apparent horizon in the region $u,v < 1$. In the region $u > 1, v <
1$ the equation for the apparent horizon is
\ben
 {\gamma \over v} - T = {1 \over 2 u} \label{eq:eighteen}
\een
If present, this is always a timelike line (since as $u$ increases, so
does $v$). For large values of $u$ the apparent horizon asymptotes to
the null line $v = {\gamma \over T}$.
It is then easy to see that the curve (\ref{eq:eighteen})
 does lie in the
region $u > 1, v < 1$ unless
\ben
 T > \gamma - \half \label{eq:nineteen}
\een
For the region $u,v >1$ the
equation for the apparent horizon is given by
\be
{1 \over 2u} + {A(v) - \gamma \over v} = 0 \label{eq:cone}
\ee
where we have defined
\be
A(v) \equiv {\beta T\sqd \over (T+(\beta \sqd - T)v^\delb)}
\label{eq:ctwo}
\ee

The singularities correspond to the zeroes of $\partial_\phi \Omega(u,v)$.
This means that $\Omega = \Omega_{cr} = \half(1 + \klog 2)$. From the
above solution it may be seen that in the region $u,v < 1$ there are no
singularities for $\beta \ne 0$. (For $\beta = 0$ the boundary coincides
with the critical line in this region as we will see later).

In the region $u > 1, v < 1$ the solution is exactly like the solution of
the RST model.  In the RST model a singularity is formed whenever the {\em
energy density} exceeds a critical value. Thus for a delta function shock
wave a spacelike singularity is always formed.  In this case, however,
there is a critical value of the {\em total energy} beyond which a
singularity is formed. For a shock wave one must have $T > T_c$, where
$T_c$ may be determined as follows. The singularity, if present must
intersect the $u=1$ line or the $v = 1$ line. It is easy to see from
(\ref{eq:seventeen} )
that there are no zeroes of $\Omega (u,v)$ along the $u = 1$
line. Along $v = 1$ one has $\Omega= \Omega_{cr}$ at a value of $u = u_0$
which is determined by solving the equation
\ben
 {u_0} = {\klog (u_0) + 2(\Phi_0 - T) \over 2(\gamma - T)}
\label{eq:twenty}
\een
where
\ben
\Phi_0 = \half(1 + \klog (2 \gamma))
\een
Note that $\half < \Phi_0 < \gamma$.  When $T > \gamma$ this always has a
solution. In this regime the singularity starts out as spacelike,
intersects the apparent horizon where it turns time-like and continues all
the way to $u = \infty$ as a timelike naked singularity, just as in the
RST model. However the solution should not be continued beyond the point
of intersection of the sinmgularity and the apparent horizon
which is given by $(u_s,v_s)$ where
\bea
u_s & = & {1 \over 2T}[e^{2T} - 1] \nn \\
v_s & = & {\gamma \over T}[1 - e^{-2T}]
\label{eq:cthree}
\eea
As noted in
\cite{RST}, $\Omega$ attains vaccum values along the line $v = v_s$.
This is the case in our model
with $\beta \ne 0$ as well. Consequently one may impose boundary
conditions corresponding to a "thunderpop" such that the solution in the
region $v > v_s$ is the linear dilaton vaccum. In our case this turns
out to be
\be
e^{-2\phi} = \kappa u ({\gamma \over v} - T)
\ee
In the region $v > v_s$ the critical line $\Omega = \Omega_{cr}$
coincides with the apparent horizon. On this line the curvature
is finite so there is no naked singularity.
One then has a picture
where the black hole has evaporated completely and information is lost in
this semiclassical model.

For $T < \gamma$, (\ref{eq:twenty})
has a solution for $T > T_c$ where $T_c$ is
value of $T$ for which the straight line represented by the left hand side
of (\ref{eq:twenty}) is tangential to the curve on
the right hand side for some
value of $u$.  This means that $T_c$ is determined by the equation
\ben
 \gamma = {T_c \over (1 - e^{-2T_c})} \label{eq:twentyone}
\een
Note that $T_c < \gamma$ as required. A plot of the critical energy $T_c$
as a function of $\beta$ is shown in Fig. 3. Note that for $T_c < T <
\gamma$ the equation (\ref{eq:twenty}) has two
solutions so that before imposition
of an additional boundary condition at $(u_s, v_s)$ the singularity meets
the outgoing pulse at a finite value of $u$. For $T > T_c$ the fate
of the singularity in this region is identical to that for $T > \gamma$
which is described above.

The interesting point is that $T_c$ is always greater $T_0 =\beta
\sqd$ which is the critical value of $T$ beyond which the boundary
starts receeding with infinite speed. A plot of the quantity ${T_c
\over T_0}$ is shown in Fig. 4. It also follows from the definition of
$\gamma$ that $\gamma - \half < \beta \sqd < T_c $.

In the region $u,v > 1$ there are no singularities for $T < \beta\sqd$.
For $\beta \sqd < T < T_c$ there is a singularity which starts
out at the boundary at $u=v=u_0$ along a timelike direction. It then
turns spacelike and meets the apparent horizon after a finite proper
time at the verge of turning timelike \footnote{We would like to
thank M. O'Laughlin for pointing
us a mistake about the nature of the singularity in this region in
an earlier version of the paper}.
This happens at the point
$(u_s,v_s)$ which is obtained by solving the following implicit
equations
\be
u_s = {v_s \over 2 \gamma \alpha}~~~A(v_s) = -\half {\rm log} (\alpha)
\ee
where $A(v)$ has been defined above in (\ref{eq:ctwo}) and $\alpha$
is the nontrivial solution of
\ben
\gamma \alpha -\half {\rm log} \alpha = \gamma
\een
As before one can impose boundary conditions here such the region
$v > v_s$ is a linear dilaton vacuum which is given in this case
by
\ben
e^{-2\phi} = \kappa \gamma \alpha {u \over v}
\een
Recall that $u_0$ is the value of $v$ for which $\xm
(v)$ blows up. Thus, just as in the classical solution, the region $v >
u_0$ is beyond the singularity and not present in the spacetime defined by
the semiclassical solution. For $T > T_c$ the singularity in the $u, v >
1$ region starts off as usual at the boundary, but crosses the reflected
wave at some finite value of $v$ where it joins the singularity of the $u
> 1, v < 1$ region as discussed above. In the Kruskal coordinates the
singularity is asymptotic to the runaway part of the boundary, similar to
that in the classical solution.

\subsection{Solutions with $\beta = 0$ }

We now mention briefly the features of the semiclassical solution for
$\beta = 0$ \footnote{This section summarizes results obtained in
collaboration with E. Martinec.}.  We will thus solve the equation
(\ref{eq:bctwelve}) with $\beta = 0$
with $T(u) \rightarrow T(u) - {1 \over 2 u^2}$
to account for the vacuum energy. It is immediately clear that in a region
where $T(u) - {1 \over 2 u^2} > 0$ the quantity $\hm^2 < 0$ which means
that $\pv \xm (v)$ is negative. This means that the physical coordinates
which are asymptotically minkowskian are double valued as a function of
$v$ and the boundary becomes spacelike. For a shock wave given by
(\ref{eq:thirteenh})
this always happens since a shock wave corresponds to an
infinite energy density. For such a shock wave the solution for $\xm(v)$
is given by
\ben
 \xm (v) = -{1 \over 2 v} + T \theta (1-v) \label{eq:bzone}
\een
Note that unlike $\beta \ne 0$ the coordinate field $\xm (v)$ is
discontinuous at the location of the pulse, with a {\em finite}
discontinuity. For a spread-out pulse this discontinuity is absent, but
corresponds to a region where $\xm$ runs backward, corresponding to the
boundary becoming spacelike.

In this case a singularity is present for any positive value of $T$. Now
one has
\bea
\Omega (u,v) & = & {u \over 2v} + T(u-1)[\theta(v-1) - \theta(u-1)] \nn
\\
& & -\half \klog({u \over 2 v}) \label{eq:bztwo}
\eea
 The boundary $u=v$ is
precisely the location of $\Omega = \Omega_{cr}$.  For the region $u > 1,
v < 1$ the singularity forms at the wake of the incoming pulse and has a
fate similar to the RST model.  In fact the case $\beta = 0$ is almost
identical to the RST model.

\subsection{Information loss}

Whenever parameters in the theory are in a range such that a singularity
and a horizon is formed there is the standard information loss at the
semiclassical level. At the semiclassical level the "in" and "out" modes
of the matter fields $f^i$ are defined as
\bea
a_\nu & = & \int du~[\xp (u)]^{i\nu} \pu f^i \nn \\
b_\nu & = & \int du~[- \xm (u)]^{-i\nu} \pu f^i\label{eq:twentyfive}
\eea
since the asymptotically Minkowskian coordinates are $\pm \klog (\pm
X^\pm)$ in all the regions. In the $\xp = u$ gauge all the nontriviality
of the mode expansions lie in the form of $\xm (u)$. If there are no
horizons $\xm (u)$ would vanish only at $u \rightarrow \infty$ and both
the integrals in (\ref{eq:twentyfive})
would be over the full range of $u$. However
in the presence of a horizon, the second integral
in (\ref{eq:twentyfive}) goes
from $-\infty$ to $-v_h$ where $v = -v_h$ is the location of the horizon.
The resulting $n_\nu$ no longer form a complete set and one has a standard
story. A reasonable definition of the $S$-matrix in the problem would be
given by
\ben
 <0,out| \prod_\nu b_\nu \prod_\nu a_\nu |0,in> \label{eq:twentysix}
\een
and would be nonunitary semiclassically.

Normally one would get a unitary evolution if one took into account the
states inside the horizon as well, i.e. if one integrates over the full
range of $v$. Here, however, the $\xm (v)$ is a nonmonotonic function of
$v$. As a result, the modes $b_\nu$ are not complete, all combinations of
the modes $a_\nu$ cannot be expressed as combinations of $b_\nu$, though
the converse is certainly true. Thus the $S$-matrix satisfies $S S^\dagger
= 1$ but $S^\dagger S \ne 1$.

\section{Quantum Fluctuations and Complementarity}

We now consider aome aspects of the quantum fluctuations of the graviton
dilaton degrees of freedom. In this model these degrees of freedom are
represented by the chiral fields $\xp (u), \xm (v), \xtm (v), \xtp (u)$.
The model is essentially a free field theory in these variables and one
may hope to address exact quantum questions.

\subsection{Light cone quantization}

We will quantize the model around the classical solution with a shock wave
described in section 2. It is sufficient to consider the set of fields
$\xp (u)$, $\xtp (u)$ and $f^i (u)$ since the boundary conditions ensure an
identical situation in the other chiral sector. We thus expand the
operators as
\bea
\xp (u) & = & \xpcl (u) +  \xpq (u) \nn \\
\xtm (u) & = & \xtmcl (u) + \xtmq (u) \nn \\
f^i (u) & = & \ficl (u) + \fiq (u) \label{eq:twentyeight}
\eea
 where subscripts
$(cl)$ and $(qu)$ denote classical and quantum pieces.  There is a
constraint
\ben
\pu \xp (u) \pu \xtm (u) + \half \pu f^i(u) \pu f^i(u) = 0
\label{eq:twentynine}
\een
In the quantum theory the right hand side has to be replaced by a suitable
normal ordering term ${1 \over 2u^2}$.  Our model is almost identical to
critical bosonic string theory. However, as emphasized in \cite{VV} the
expansion of the field in terms of the physical modes are rather
different.

We will quantize the theory in a light cone gauge $\xp (u) = \pp u$.  We
will also put the system in a box of size $2 \pi \tL$ to keep track of
infrared behaviour. Using standard mode expansions
\ben
\pu f^i_{qu}(u) = {1 \over \spi \tL}\sum_{m}\tf^i_m~e^{-im{u \over \tL}}
\label{eq:thirty}
\een
(and similarly for the other fields) one may then solve the constraints to
solve for the nonzero modes of $\xtm (u)$ in terms of the $\tf^i_m$.  The
remaining dynamical variables are the canonically conjugate pair of zero
modes $q^-,\pp$ and the matter oscillators. The zero mode $q^-$ represents
the overall translational degree of freedom and decouples from the
dynamics. At the classical level this is the arbitrary integration
constant which is chosen to ensure that $\pm \klog[\pm X^\pm]$ are good
asymptotic coordinates.

The mode expansions of the type (\ref{eq:thirty})
are, however, not very useful in
this problem. In the quantization around the specific solution
representing the formation and evaporation of black hole, the asymptotic
"in" coordinates are $ \yp = \klog u$ rather than $u$. Thus the
oscillators $\tf^i_m$ do not annahilate the "in" vacuum. In the region $u
> 0$ one should thus expand in terms of the appropriate modes
\ben
\pyp f^i_{qu}(\yp) = {1 \over \spi
L}\sum_{m} f^i_m~e^{-im\ypl}\label{eq:thirtyone}
\een
 In the region $u < 0$
one has a similar modes in terms of coordinates $\klog (-u)$. We will also
introduce coordinates $\ym = \klog(v)$.  The semiclassical region of
interest has $v < 1$ and hence $\ym < 0$.  The boundary in terms of these
coordinates is $\yp = \ym$.

In the following we shall deal exclusively with the region $u > 0$.
However to examine questions like unitarity etc. one has to deal with the
oscillators corresponding to the other region as well \footnote{In the
covariant quantization in \cite{VV} one has a similar doubling of states
corresponding to the two signs of the zero modes of $xp$. The vacuum
sector S-matrix is unitary only if the full space of states are taken into
account.}.

Let us now solve the constraints in this gauge. The classical solution for
$\xtm(u)$ is given by
\ben
\xtm_{cl} (u) = \beta^2 a \theta (-\yp) \label{eq:thirtytwo}
\een
while $f^i_{cl}$ is nonzero only along $\yp =0$. Then the quantum part of
$\xtm (u)$ may be solved to be, for $\yp > 0$
\ben
 \pyp \xtm (\yp) = {e^{-\yp} \over 2 \pi L^2} \sum_m \ltr_m e^{-im\ypl}
\label{eq:thirtythree }
\een
where $\ltr$ denotes the virasoro modes of the matter energy momentum
tensor
\ben
\ltr_m = \half[\sum_n : f_{m-n} f_n : - 2\pi L^2]\label{eq:thirtyfour}
\een
and we have included the standard intercept coming from normal ordering.
The "in" vacuum is now defined by
\ben
 f_n \keto = 0 ~~~~~~n \geq 0 \label{eq:thirtysix}
\een
The Virasoro generators $\ltr_m$ act on the vacuum as follows
\bea
 \ltr_m \keto & = & 0~~~~~~m > 0 \nn \\
\ltr_0 \keto & = & - 2\pi L^2 \keto
\label{eq:thirtyseven}
\eea

\subsection{Fluctuations of the boundary}

In terms of the coordinates $\yp, \ym$ the invariant line element on the
boundary is given by $ds_{B}$ where
\ben
 ds_{B}^2 = e^{2\phi_0} [\pyp \xp (\yp) \pym \xm (\ym)]_{\yp = \ym}
\label{eq:thirtyeight}
\een
where $\phi_0$ is the constant value of the dilaton field at the
boundary. (Recall that the dilaton field is maintained to be
a constant along the boundary in the quantum theory so that there
are no quantum fluctuations of this quantity). For the classical
solution we are expanding around, $e^{2\phi_0} = {1 \over \beta^2}$.
We will now compute the dispersion of the quantity $ds_{B}^2$ along the
boundary in the region $\yp, \ym < 0$.

In a quantum theory of gravity correlations of local operators do not make
much sense, since one is integrating over the metric. The meaning of the
dispersions we are about to calculate is as follows. We are expanding
around a specific classical solution corresponding to a dynamical black
hole formation with some incoming matter. We are interested in getting an
idea of the strength of quantum fluctuations of space-time quantities, one
example of which is the line element on the boundary. This may be done by
computing the corresponding local quantities {\em using the classical
metric as the reference metric}. These quantities have a perfectly good
meaning for an asymptotic observer. In the asymptotic region the
fluctuations of the metric are weak and the asymptotic observer may thus
use the classical metric to make measurements. The correlations of these
local operators have physical meaning {\em only} for these asymptotic
observers.

The correlations of $ds_{B}$ require mode expansions for $\pym \xm (\ym)$.
Unlike $\pyp \xtm$ these are difficult to compute exactly. We shall obtain
these for small fluctuations around the classical solution. It is most
convenient to consider the equation (\ref{eq:bctwelve})
in appropriate coordinates,
as the full quantum equation and expand this around the classical
solution. To lowest order one gets in the region $\ym < 0$, where
$\hm^{cl} = \beta e^{-\ym}$,
\ben
\beta[\pym \hm^q + 2 \hm^q] = e^{-\ym}[1 - \half \pym f_{qu} \pym
f_{qu}] \label{eq:thritynine}
\een
 Here $\hm^q$ denotes the quantum part of
$\hm$. Defining modes as
\ben
\hm^q (\ym) = {e^{-\ym} \over 2\pi L^2}\sum_m h_m e^{-im\yml}
\label{eq:forty}
\een
one has to this order
\ben
 h_m = {\ltr_m \over \beta(i{m \over L} - 1 )}\label{eq:fortyone}
\een
 From this expression one may obtain the modes of $\pym \xm (\ym)$.  The
correlator
\ben
 \brao ds_{B}^2(\ym) ds_{B}^2 (\ymp) \keto
\een
may be now calculated using the Virasoro algebra. For $\ym \rightarrow
\ymp$ this goes as
\ben
 \brao ds_{B}^2(\ym) ds_{B}^2 (\ymp) \keto \sim {1 \over
\beta^2 (\ym - \ymp)^2}
\label{eq:fortytwo}
\een

A freely falling observer will make measurements with a cutoff which keeps
the invariant local distance fixed to some value and would replace the
right hand side with
\ben
 {1 \over (\ym - \ymp)^2 + b^2 e^{-2\rho_{cl}[\half(\ym + \ymp)]}}
\een
where $\rho_{cl}$ is the conformal mode of the classical metric. Here $b$
is the geodesic cutoff. The quantity $e^{-2\rho_{cl} (\ym)}$ along the
boundary is small near the singularity, but not near the horizon. Thus the
freely falling observer perceives boundary fluctuations which are large
only at scales smaller than the cutoff $b$. In particular he/she does not
perceive any strong fluctuation near the horizon. This is consistent with
the fact that the freely falling observer does not see anything special at
the horizon.

An asymptotic observer at future null infinity would make measurements
with a cutoff which has some fixed value in terms of the aymptotic
coordinates. Recall that the asymptotic coordinates in the classical
solution are
\ben
w^- \equiv -\klog[-\xm(v)] = -\klog[e^{-\ym} - a] \label{eq:fortythree}
\een
Since
\ben
\delta \ym = \delta w^- [ 1 - ae^{\ym} ]
\een
the quantity (\ref{eq:fortytwo}) is
\ben
 \brao ds_{B}^2(\ym) ds_{B}^2 (\ymp) \keto \sim {1 \over
\beta^2 (\delta w^-)^2
[1-ae^{\ym}]^2 }\label{eq:fortyfour}
\een
 The quantum dispersion at scales
larger than the cutoff of the aymptotic observer, ${\tilde b}$ is obtained
by setting $\delta w^- = {\tilde b}$.

For $a < 1$ the fluctuations of the boundary increase as one gets closer
to the point of impact at $\ym = 0$, but never become infinite. For $a >
1$ the fluctuations become strong at the location of the horizon
regardless of the value of the cutoff. We thus conclude that as the
incident energy approaches the critical value and a horizon starts to
form, the asymptotic obsever measures very large fluctuations near the
horizon. The basic reason behind this phenomenon is the large redshift
between the horizon and infinity. Our result is in fact a concrete
illustration of the contention of 't Hooft.

Quantum dispersions of other quantities show a similar behaviour. For
example the correlation of the rescaled curvature ${\tilde R}$ introduced
in (\ref{eq:bcdtwo}) may be calculated to be (in the region $u >1, v< 1$)
\bea
 \brao & & \curv (\yp,\ym) \curv (\ypp,\ymp) \keto = \nn \\
& & (1 + e^{(\ym-\yp)})^2 [{1 \over (\delta \yp)^2} + {1 \over (\delta
\ym)^2}]\label{eq:fortyfive}
\eea
 By the same reasoning as above, curvature
fluctuations as measured by the aymptotic observer grow very large near
the horizon.

\section{In lieu of a conclusion}

The real question is : does these large fluctuations of space-time as
perceived by the asymptotic observer invalidate the semiclassical
approximation regardless of the mass of the black hole ? The question is
confused by the fact that these fluctuations are large only for an
asymptotic observer and one might argue that to examine the validity of
the semiclassical approximation one should use a local geodesic cutoff. It
is nevertheless clear that the asymptotic observer will have to account
for large space-time fluctuations in his or her description of the Hawking
process.

One way to obtain a definitive answer to this important question is to
calculate quantities which have unambiguous meaning in quantum gravity,
for example correlations of integrated operators. These "susceptibilities"
may be studied as a function of the parameter $a$ of the classical
solution and one may be able to see whether quantum corrections grow large
as $a$ approaches the critical value $1$. In this regard the present model
with nonzero $\beta$ could be useful since this has a critical value of
$a$ necessary for black hole formation and one could study quantum
properties of the model as $a \rightarrow 1$ from below.

\acknowledgements

We would like to thank E. Martinec for discussions and collaboration
during the initial stages of this work. We also thank T. Banks, P. Joshi,
G. Mandal, S. Mathur, M. O'Laughlin, A. Sen,  T.P. Singh, S. Shenker and
S. Wadia for discussions during various stages of the work. We are
especially grateful to Sandip Trivedi for very fruitful discussions.
S.R.D. would like to thank members of the Theory Group of Enrico Fermi
Institute, University of Chicago and the Theory Group at Rutgers
University for hospitality during the early stages of this work.
One of the authors (S.M.) would like to thank Professor Abdus Salam,
the International Atomic Energy Agency and UNESCO for hospitality
at the International Center for Theoretical Physics, Trieste.

\figure{Fig. 1.: Classical Solution for $a < 1$. The incident shock
wave reflects off the boundary and proceeds to future null
infinity. No singularities are formed \label{one}}
\figure{Fig. 2.: Classicial solution for $a > 1$. The boundary runs
away for a value of $u$ which is {\em greater} than the
value of $u$ at which the shock wave hits the boundary. A
singularity is formed which is null-like in the region above
the reflected wave and space-like in the region below the
reflected wave. The space-like portion of the singularity
asymptotes to the global event horizon \label{two}}
\figure{Fig.3.: The critical energy $T_c$ beyond which there is a
singularity in region below the reflected wave plotted as
a function of $\beta$ \label{three}}
\figure{Fig.4. : The ratio of $T_c$ to the minimum energy for black
hole formation $T_0$ as a function of $\beta$ \label{four}}
\newpage
\begin{center}
List of changes
\end{center}

The Editor
Physical Review D

                   Re: Manuscript DP5003

Dear Sir,

We are sending separately a REVTEX version of our paper entitled
"Black Hole Formation and Space-time fluctuations in two dimensional
dilton gravity and complementarity" (Manuscript No. DP5003) which
has been accepted fot publication in Physical Review. The
figures are sent separately as an unencoded compressed tar file.

As suggested by you in an earlier communication, we are detailing
below the minor changes we have made in the manuscript. In
the following the equations refer to the version of the manuscript
which we are sending now. Of course, these modifications do not
change the main content and conclusions of the paper.

1. The equation (1) has been corrected for some signs and $e^{-\phi}$
has been replaced by $e^{-2\phi}$.

2. At the end of the paragraph following equation (1) we had added
the definitions of $u$ and $v$ used in the paper.

3. A typing mistake in equation (26) has been corrected. In the
second line of the equation we have replaced $u > 0$ by $u > 1$.

4. A mistake in equation (27) has been corrected (there should be
an overall factor of ${1 \over \kappa}$ which was missed earlier.

5. An overall factor of $\kappa$ which was missed in (28) has been
suppplied.

6. In equations (30) and (31) ${\hat \kappa}$ has been replaced
by ${{\hat \kappa} \over \kappa}$.

7. A factor of half was missing in equation (32) which has now
been inserted.

8. In equation (34) $Tu\theta(u-1)$ was wrongly typed as
$Tu\theta(1-u)$ earlier. This has been corrected.

9. In equation (36) $v^{{\sqrt{\Delta}}\over \beta}$ was typed
as $u^{{\sqrt{\Delta}}\over \beta}$. This has been corrected.

10. In the line following equation (38) we have added a line
which states the asymptotic behaviour of the apparent horizon.

11. After equation (39) we have given the equation for the
apparent horizon in the region $u,v > 1$. This is
equation (39-40) and have been added for the
sake of completeness.

12. In the paragraph after equation (43) we have added another
equation (eqn. 44) which gives the values of $u$ and $v$ where the
apparent horizon meets the singularity.

13. At the end of the same paragraph we have added, for completeness,
the form of the dilaton field in the region $v > v_s$in eqn. (45).  We
have also added a line which states the behaviour of the critical line
in this region.

14. In the last paragraph of section IV A, we have added three
equations, two (equations 47 and 48) denoting the point of
intersection of the apparent horizon and the singularity and the other
(eqn. (49)) which gives the dilaton field in the region above this
intersection points.

15. In equation (63) a factor of $e^{2\phi_0}$ was missing which
has been inserted.

16. In equation (66) a typing mistake has been corrected. "2" should
be replaced by "1".

17. In equation (68) a factor of ${1 \over \beta^2}$ was missing.
This has been now inserted.

18. Some of tyhe references have been updated (journal references
have been given wherever possible.)

19. We have added Figure captions, as required by the submission
guidelines.

20. We have modified the labels of Figs. 3 and 4 since the
phrase "critical energy" could be a bit misleading since there
are several critical values of energy in the problem. We have made
them precise using the notation used in the text.

We shall appreciate if you could let us know the publication date of
the paper.

Thank you.

Yours sincerely,

Sumit Ranjan Das      Sudipta Mukherji


\begin{references}
\bbb{INFO} S.W.  Hawking, {\em Comm. Math. Phys.} {\bf 43},199
(1975);\\  \cmp, {\bf 87},395 (1982)

\bbb{REV} For
recent reviews and references, see e.g. J. Harvey and A. Strominger,
in {\em String Theory and Quantum Gravity '92} ed. J. Harvey {\em et.
al.} (World Scientific, 1994);
S. Giddings, hep-th/9209113 and hep-th/9306041; J.
Preskill, hep-th/9209058; D. Page, hep-th/9305040.

\bbb{MSW} G. Mandal, A. Sengupta and S.R. Wadia,
{\em Modern} \\  {\em Physics} {\em Letters} {\bf A6}
,1685 (1991)

\bbb{WIT} E. Witten, \pr {\bf D44}, 314 (1991)

\bbb{CGHS} C. Callan, S. Giddings, J. Harvey and A. Strominger, \pr
{\bf D45} ,R1005 (1992)

\bbb{BANKS}T. Banks, A.
Dabholkar, M.  Douglas and M. O'Laughlin, \pr {\bf D45},3607 (1992)

\bbb{RST} J. Russo,
L. Susskind and L. Thorlacius, {\em Phys. Lett} {\bf B292}
, 13 (1992); L.
Susskind and L.  Thorlacius, \nph {\bf B383}, 123 (1992);
J. Russo, L. Susskind
and L. Thorlacius,
\pr {\bf D46} (1992) 3444; \pr {\bf D47} (1993) 533.

\bbb{BGHS}B. Birnir, S. Giddings, J. Harvey
and A. Strominger, \pr {\bf D46} (1992) 638.

\bbb{HAW} S.W. Hawking, {\em Phys. Rev.Lett.} {\bf 69},406 (1992);
S.W. Hawking and J.M.Stewart, {\em Nuclear Physics}, {\bf B400}
,393 (1993).

\bbb{STROM}A. Strominger, \pr {\bf D46} ,4396 (1992)

\bbb{BILAL} A. Bilal and C. Callan, \nph, {\bf B394},73 (1993).

\bbb{DEALWIS} S. de Alwis, {\em Phys. Lett.} {\bf B289},278
(1992), {\em Phys. Lett.} B300,330 (1993), \pr {\bf D46},5429 (1992)

\bbb{GIDST} S. Giddings and A. Strominger, \pr {\bf D47}, 2454 (1993).

\bbb{PARK} Y. Park and A. Strominger, \pr
{\bf D47}, 1569 (1993).

\bbb{BANK} T. Banks, M. O'Laughlin and A. Strominger, \pr
{\bf D47},4476 (1993).

\bbb{LOWE} D. Lowe, \pr {\bf D47},2446 (1993).

\bbb{PIRAN} T. Piran and A. Strominger, \pr {\bf D48}, 4729
(1993).

\bbb{MARTIN} D. Lowe and M. O'Loughlin, \pr {\bf D48}, 3735
(1993).

\bbb{TRIV} S. Trivedi, \pr {\bf D47},4233 (1993);
A. Strominger and S. Trivedi, \pr {\bf D48}, 5778 (1993).

\bbb{BHMM} S.R. Das, \mpla,{\bf A8},69 (1993); S.R. Das in
{\em String Theory and Quantum Gravity '92} ed. J. Harvey {\em et.
al.} (World Scientific, 1994);
E. Martinec and S. Shatashvilli, \nph, {\bf B368}, 338 (1992)

\bbb{DMWBH} A. Dhar, G. Mandal and S.R. Wadia, \mpla,
{\bf A7},3703 (1992).

\bbb{WAVE} S.R. Das,
\mpla, {\bf A8}, 1331 (1993); A. Dhar, G. Mandal and S.R. Wadia, \mpla,
{\bf A8},1701 (1993).

\bbb{JEVYON} A. Jevicki and T. Yoneya, {\em Nuclear Physics}~
{\bf B411}, 64 (1994).

\bbb{RUSSOA} J.G. Russo, {\em Phys. Lett.} {\bf B300},336 (1993).

\bbb{VV} H. Verlinde and E. Verlinde, \nph, {\bf B406},43 (1993);
K. Schoutens, H. Verlinde and E. Verlinde, \pr ,{\bf D48},2690 (1993).

\bbb{MIRR} See N. Birrell
and P. Davies, {\em Quantum Fields in Curved Space} (Cambridge,1982) and
references therein; R. Carlitz and S. Willey, \pr,{\bf D36}
2327 (1987); F. Wilczek, hep-th/9302096.

\bbb{HVER} T. Chung and H. Verlinde, hep-th/9311007.

\bbb{THOFT} G. 't Hooft, \nph, {\bf B335}, 138 (1990);\\
{\em Physica Scripta}, T36, 247 (1991) and
references therein.

\bbb{SUSSTH} L. Susskind,
L. Thorlacius and J. Uglum, \\ hep-th/9306069;
L. Susskind and L. Thorlacius, \pr {\bf D49}, 966 (1994).

\bbb{SUSS}  L. Susskind, {\em Physical Review Letters} {\bf 71},
2367 (1993); L. Susskind, hep-th/9308139.

\bbb{TRP} S. Trivedi, {\em private communication}
\end{references}
\end{document}